\documentclass[runningheads,a4paper]{llncs}

\usepackage{amssymb}
\usepackage{graphicx}
\usepackage{epstopdf}

\usepackage{url}

\begin{document}

\mainmatter  

\title{Challenges in Kurdish Text Processing}
\vspace{-2cm}


\author{Kyumars Sheykh Esmaili
}
%

\institute{Nanyang Technological University\\
N4-B2a-02, 639798 Singapore\\
\url{ kyumarss@ntu.edu.sg}  \\
}

\maketitle

\vspace{-0.5cm}
\begin{abstract}
Despite having a large number of speakers, the Kurdish language is among the less-resourced languages. In this work we highlight the challenges and problems in providing the required tools and techniques for processing texts written in Kurdish. From a high-level perspective, the main challenges are: the inherent diversity  of the  language, standardization and segmentation issues, and the lack of language resources. 

\end{abstract}

\section{Introduction}
\vspace{-0.1cm}
Kurdish language belongs to the Indo-Iranian family of Indo-European languages. Its closest better-known relative is Persian. Kurdish is spoken in Kurdistan, a large geographical area spanning the intersections of  Iran, Iraq, Turkey, and Syria.  It is one of the two official languages in Iraq and has a regional status in Iran.

Despite having 20 to 30 millions of speakers\footnote{Numbers vary, depending on the source.}, Kurdish is among the less-resourced languages and there are very few tailor-made tools available for processing texts written in this language.
Similarly, it has not seen much attention from the IR and NLP research communities and the existing literature can be summarized as a few attempts in building corpus and lexicon for Kurdish~\cite{techChal,soralex}. 
In order to provide the basic tools and techniques for processing Kurdish, we have recently launched a project at University of Kurdistan (UoK \footnote{Project's website: http://eng.uok.ac.ir/esmaili/research/klpp/en/main.htm.}). This paper gives an overview of the main challenges that we need to address throughout this project.

Before proceeding to enumerate the challenges, we would first like to highlight a few things about the scope and methodology of the current paper. Firstly, in this work we only consider the two largest and closely-related branches of Kurdish --namely Kurmanji (or Northern Kurdish) and Sorani (or Central Kurdish)--  and exclude the other smaller and distant dialects.  Secondly, in the interest of space,  we give greater importance to the issues that are specific to Kurdish and refer to the related papers for  in-depth discussion of issues that are shared between Kurdish and other languages (in particular, Persian, Arabic, and Urdu).  Finally, we restrict our discussion to the  bag-of-words (i.e., IR) model and do not address the  structural (i.e., NLP) aspects.

\section{Challenges}
\vspace{-0.1cm}
We have categorized the main challenges into five groups. While the first two groups are concerned with the diversity aspect of the Kurdish language, the third and fourth highlight the processing difficulties and the last one examines the depth of resource-scarcity for Kurdish.

\vspace{-0.2cm}
\ \\
\textbf{2.1~~~Dialect Diversity}\\
The first and foremost challenge in processing Kurdish texts is its dialect diversity. In this paper we focus on Kurmanji and Sorani which are the two most important Kurdish dialects in terms of number of speakers and degree of standardization~\cite{kurdishOverview}. Together, they account for more than 75\% of native Kurdish speakers~\cite{soralex}.

The features distinguishing these two dialects are mainly morphological (the phonological differences are explained in the next section). The important morphological differences are~\cite{mackenzie,kurdishOverview}:
\begin{itemize}
	\item  Kurmanji is more conservative in retaining both gender (feminine:masculine) and case opposition (absolute:oblique) for nouns and pronouns. Sorani has largely abandoned this system and uses the pronominal suffixes to take over the functions of the cases\footnote{Although there is evidence of gender distinctions weakening in some varieties of Kurmanji~\cite{kurdishOverview}.}. 
	\item  in the past-tense transitive verbs, Kurmanji has the full ergative alignment but Sorani, having lost the oblique pronouns, resorts to pronominal enclitics.
	\item  in Sorani, passive and causative can be created exclusively via verb morphology, in Kurmanji they can also be formed with the verbs \texttt{hatin} (to come) and \texttt{dan} (to give) respectively. 
	\item  the definite suffix \texttt{-eke} appears only in Sorani
\end{itemize}

\begin{figure}[ht]%
\includegraphics[width=\columnwidth]{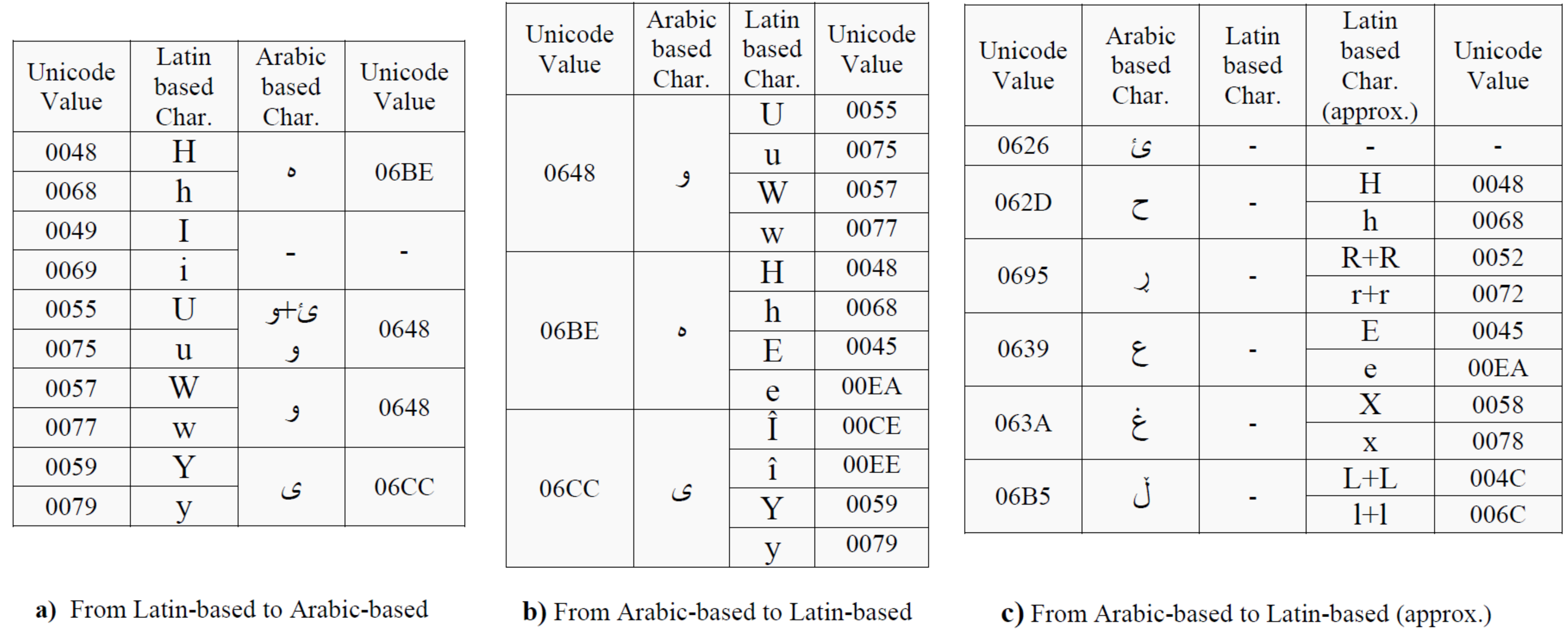}%
\vspace{-0.5cm}
\caption{\scriptsize{Non-Trivial Mappings between Arabic-based and Latin-based Kurdish Alphabets.}}%
\label{fig:mapping}%
\end{figure}

\vspace{-0.2cm}
\ \\
\textbf{2.2~~~Script Diversity}\\
Due to geopolitical reasons, each of the two aforementioned dialects  has been using its own writing system. In fact, Kurdish is considered a \textit{bi-standard} language~\cite{techChal}, with Kurmanji written in Latin-based letters and Sorani written in Arabic-based letters. Both of these systems are almost totally phonetic~\cite{techChal}.
As noted before, Sorani and Kurmanji are not morphologically identical and since these systems reflect the phonology of their corresponding dialects, there is no bijective mapping between them. In Figure~\ref{fig:mapping} we have included three tables to demonstrate the non-trivial mappings between these two writing systems. It should be noted that the table in c) contains approximate equivalences. 
These mappings are in line with the list proposed in~\cite{hassanpour}.

\vspace{-0.2cm}
\ \\
\textbf{2.3~~~Normalization}\\
The Unicode assignments of the Arabic-based Kurdish alphabet has two potential sources of ambiguity which should be dealt with carefully:  
\begin{itemize}
	\item  for some letters such as \texttt{ye} and \texttt{ka} there are more than one Unicode~\cite{persianChal}.
During the normalization phase, the occurrences of these multi-code letters should be unified.
	\item as in Urdu, the isolated and final forms of the Arabic letter \texttt{ha} constitute one letter
(pronounced e), whereas the initial and medial forms of the same
Arabic letter constitute another letter (pronounced h), for which a
different Unicode encoding is available~\cite{soralex,techChal}. In many electronic texts, these letters are written using
only the \texttt{ha}, differentiated by using the \textit{zero-width non-joiner} character that prevents a character from being joined to its
follower. This distinction must be taken into account in the normalization phase. 
\end{itemize}

\vspace{-0.2cm}
\ \\
\textbf{2.4~~~Segmentation and Tokenization}\\
Segmentation refers to the process of recognizing boundaries of text constituents, including sentences, phrases and words.
Compared to Persian and Arabic, this process is relatively easier in Kurdish, mainly because short vowels are explicitly represented in the Kurdish writing systems.

In fact, as discussed in~\cite{arabicChal}, the absence of short vowels contributes most significantly to ambiguity in Arabic language, causing difficulty in homograph resolution, word sense disambiguation, part-of-speech detection. In Persian, its negative consequence is more visible in detecting the Izafe constructs~\cite{persianChal} \footnote{Izafe is  an unstressed vocal \texttt{-e} or \texttt{-i}  added between prepositions, nouns and adjectives in a phrase. It approximately corresponds to the English preposition \textit{of}.  This construct is frequently used in both Persian and Kurdish languages.}.


Despite incorporating short vowels, the Arabic-based Kurdish alphabet still suffers from two problems which are inherited from the Arabic writing system: 
\begin{itemize}
	\item Arabic alphabet does not have capitalization and therefore it is more difficult to  recognize sentence boundaries as well as recognizing Named Entities.
	\item Space is not a deterministic delimiter and boundary sign~\cite{persianChal}. It may appear within a word or between words, or may be absent between some sequential words. There are some proposals on how to tackle this issue in Persian~\cite{step1} and Urdu~\cite{urduChal}.

\end{itemize}

\vspace{-0.2cm}
\ \\
\textbf{2.5~~~Lack of Language Resources}\\
Kurdish is a resource-scarce language for which the only linguistic resource available on the Web is raw text~\cite{soralex}. 

More concretely, in spite the few attempts in building corpus~\cite{techChal} and lexicon~\cite{soralex}, Kurdish still does not have any large-scale and reliable general/domain-specific corpus.
Furthermore, no test collection --which is essential in evaluation of Information Retrieval systems--  or stemming algorithm has been developed for Kurdish so far.

Lastly, although Kurdish is well served with dictionaries~\cite{kurdishOverview}, it still lacks a WordNet-like semantic lexicon.

\section{Conclusions}
\vspace{-0.1cm}
Kurdish text processing poses a range of challenges. The most important one is the dialect/script diversity  which has resulted in a \textit{bi-standard} situation. As the examples in~\cite{techChal} show, the ``same'' word, when going from Kurmanji to Sorani, may at the same time go through several levels of change: writing systems, phonology, morphology, and sometimes semantics. This clearly shows that the mapping between this two dialects are more than \textit{transliteration}, though less complicated than \textit{translation}. Any text processing system designed for Kurdish language should develop and exploit a mapping between these two standards.

On the technical side, providing the required processing techniques --through leveraging the existing techniques or designing new ones if needed-- offers many avenues for future work. However, as a critical prerequisite to most of these tasks, a core set of language resources must be available first. At UoK, we have taken the first step and are currently working on building a large standard test collection for Kurdish language.

\bibliographystyle{plain}      
\bibliography{refs}

\begin{thebibliography}{1}

\bibitem{arabicChal}
A.~Farghaly and K.~F. Shaalan.
\newblock {Arabic Natural Language Processing: Challenges and Solutions}.
\newblock {\em ACM Trans. on Asian Lang. Info. Processing}, 8(4), 2009.

\bibitem{techChal}
G{\'e}rard Gautier.
\newblock {Building a Kurdish Language Corpus: An Overview of the Technical
  Problems}.
\newblock In {\em Proceedings of ICEMCO}, 1998.

\bibitem{kurdishOverview}
Goeffrey Haig and Yaron Matras.
\newblock {Kurdish Linguistics: A Brief Overview}.
\newblock {\em Sprachtypologie und Universalienforschung / Language Typology
  and Universals}, 55(1), 2002.

\bibitem{hassanpour}
Amir Hassanpour.
\newblock {\em Nationalism and Language in Kurdistan, 1918-1985}.
\newblock Mellen Research University Press, 1992.

\bibitem{mackenzie}
David~N. MacKenzie.
\newblock {\em Kurdish Dialect Studies}.
\newblock Oxford University Press, 1961.

\bibitem{step1}
M.~Shamsfard, H.Sadat Jafari, and M.~Ilbeygi.
\newblock {STeP-1: A Set of Fundamental Tools for Persian Text Processing}.
\newblock In {\em Proceedings of the Seventh Conference on International
  Language Resources and Evaluation (LREC'10)}, 2010.

\bibitem{persianChal}
Mehrnoush Shamsfard.
\newblock {Challenges and Open Problems in Persian Text Processing}.
\newblock In {\em Proceedings of the 5th Language and Technology Conference
  (LTC)}, 2011.

\bibitem{soralex}
G{\'e}raldine Walther and Beno{\^\i}t Sagot.
\newblock {Developing a Large-scale Lexicon for a Less-Resourced Language}.
\newblock In {\em {SaLTMiL's Workshop on Less-resourced Languages (LREC)}},
  2010.

\bibitem{urduChal}
U.~I.~Bajwa Z.~Rehman, W.~Anwar.
\newblock {Challenges in Urdu Text Tokenization and Sentence Boundary
  Disambiguation}.
\newblock In {\em Proceedings of the 2nd Workshop on South Southeast Asian
  Natural Language Processing (WSSANLP)}, 2011.

\end{thebibliography}

\end{document}